\documentclass{elsart}

 \usepackage{graphics}
 \usepackage{epsfig}

\usepackage{amssymb}

\begin{document}

\begin{frontmatter}



\title{Symbiosis in the Bak-Sneppen Model for Biological
Evolution with Economic Applications}

\author{M. Bartolozzi\thanksref{a}\corauthref{*}}
\ead{mbartolo@physics.adelaide.edu.au}
\author{, D. B. Leinweber\thanksref{a}}
\author{, A. W. Thomas\thanksref{a}\thanksref{b}}
\address[a]{Special Research Centre for the Subatomic 
Structure of Matter (CSSM), University of Adelaide, Adelaide, SA 5005,
Australia}
\address[b]{Jefferson Laboratory, 12000 Jefferson Ave., Newport News,
  VA 23606, USA}
\corauth[*]{Corresponding author: CSSM, Rm. 126, Lvl 1 Physics Building,
 University of Adelaide, SA 5005, Australia.}

\begin{abstract}
 In the present work we extend the Bak-Sneppen  model for biological evolution by
introducing local interactions between species.  This
``environmental'' perturbation modifies the intrinsic fitness of
each element of the ecology, leading to higher survival probability, 
even for the less fit.  While the system still self-organizes toward a
critical state, the distribution of fitness broadens, losing the
classical step-function shape.  A possible application in economics 
is discussed, where firms are represented like
evolving species linked by mutual interests.
\end{abstract}

\begin{keyword}
Complex Systems \sep Evolution/Extinction \sep Self-Organized Criticality 
\sep Econophysics
\PACS 05.65.+b \sep 89.75.Fb \sep 45.70.Ht \sep 89.65.Gh
\end{keyword}
\end{frontmatter}

 In the past two decades several studies have been devoted to the 
investigation of the ubiquitous presence of power laws in natural and
social systems.  An important contribution to this field of research
has been given by Bak, Tang and Wiesenfeld (BTW)~\cite{Bak87,Bak88}, who 
developed  the concept of self-organized criticality (SOC).  The key
idea behind SOC is that complex systems, that is systems constituted
of many interacting elements, although obeying different microscopic
physics, may exhibit similar dynamical behaviour, statistically
described by the appearance of power laws in the distributions of
their characteristic features.  The lack of a characteristic scale,
indicated by the power laws, is equivalent to those of physical
systems during a  phase transition -- that is at the critical point.
It is worth emphasizing that the original idea~\cite{Bak87,Bak88}
was that the critical state is reached ``naturally'', without 
any external tuning. This is the origin of the adjective {\em self-}organized.
 In reality a certain degree of tuning is
necessary: implicit tunings like local conservation laws 
and specific boundary
conditions seem to be important ingredients for the appearance of 
power laws~\cite{Jensen}.
   
The classical example of a system exhibiting SOC behaviour  is the 2D
sandpile model~\cite{Bak87,Bak88,Jensen,Turcotte99}. Here the cells of a
grid are randomly filled, by an external  driver, with ``sand''.
When the gradient between two  adjacent cells exceeds a certain
threshold a redistribution of the sand occurs, leading to more
instabilities and further redistributions.  The avalanche dynamics
that drives the system from one metastable state to another is the
benchmark of all systems exhibiting SOC.  In particular, the
distribution of the avalanche sizes, their duration and the energy
released, all obey power laws.
 
The framework of self-organized criticality  has been claimed to play
an important role in solar flaring~\cite{lu91,lu93}, space
plasmas~\cite{chang_book,valdiva_book} and
earthquakes~\cite{bak89,sornette89,sornette90,gould93,huang98}  in the
context of both astrophysics and geophysics.  In biology SOC has been
linked to the punctuated equilibrium~\cite{gould93} in species
evolution~\cite{Bak93}.  Some work has also been carried out in the
social sciences.  In particular, traffic flow and traffic
jams~\cite{nagel93,nagel95,nagatani95,nagatani96},
wars~\cite{Roberts98}, as well as
stock-market~\cite{Turcotte99,Bak93b,Bak97,Feigenbaum03} dynamics,
have  been studied.  A more detailed list of subjects and references
related to SOC can  be found in the review paper of
Turcotte~\cite{Turcotte99}.

In the present work we extend the Bak-Sneppen (BS) model for
evolution~\cite{Bak93} by introducing 
explicit coupling terms in the fitness of each species of the ecology.
We find that the equilibrium configuration of the model can be deeply influenced  
by the environmental forces, leading to a wider survival probability also
for species with a lower degree of adaptation.   
A possible application of our extension of the BS model to the economic world is that
 the distribution of global-fitness can be related to the size
distribution of firms in the most developed markets.  In this respect
the evolution of firms is seen as a punctuated equilibrium process in
which the convolution of mutual interest can justify the spreading in
size of the firms themselves.

The toy model proposed in 1993 by Bak and Sneppen~\cite{Bak93} is 
one of the most popular models for biological evolution. The
main idea behind this  model is that each species can be uniquely
characterized by a single parameter called {\em fitness}. The fitness
of a species represents its degree of adaptation with respect to the
external environment.  Highly adapted species will hardly undergo any
successful, 
spontaneous mutations.  At the opposite end of the scale, 
if a species has a very low  degree of 
fitness it needs to mutate in order to survive and its mutation
automatically influences the other species belonging to the same
environment.
These concepts can be easily formulated as a simple 1D
model. Suppose that the ecology can be represented by a periodic array
of $N$ cells and each cell, $i$,  is assigned a fitness, $B(i)$, taken from a uniform
distribution between 0 and 1.  Once we have fixed the initial
condition, for each discrete time-step,
 the dynamical evolution of the  system works as follows:

a)  locate the species with minimum fitness -- that is, the one most 
likely to mutate, $k$,

b) change the fitness of $k$ and that of its neighbours
(species related) according to
\begin{eqnarray}
 B(k-1)\rightarrow u_{1}, \nonumber \\  B(k) \rightarrow
u_{2},\nonumber \\ B(k+1)\rightarrow u_{3},
\label{mutation}
\end{eqnarray} 
where the new fitness value, $u_{i}$, is a random number taken  from a
uniform distribution bounded between 0 and 1.

\begin{figure}
\centerline{\epsfig{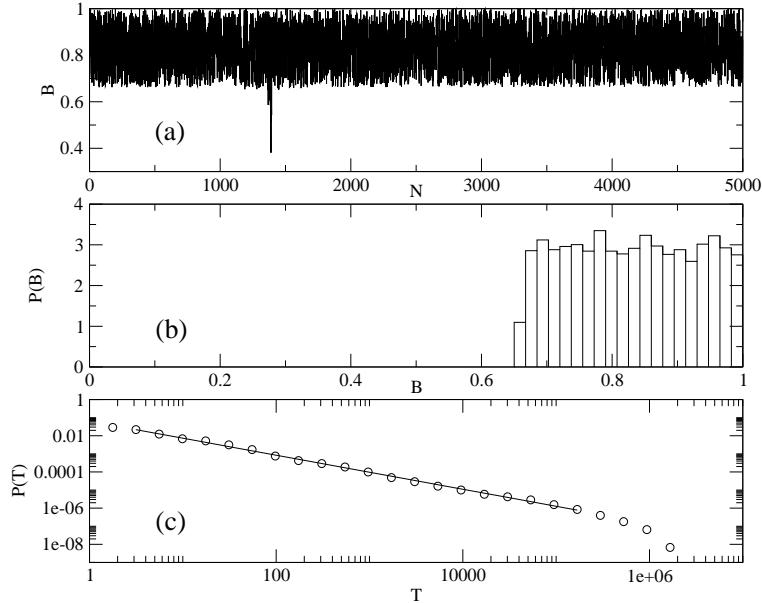}}
\caption{(a): Snapshot of the fitness, $B$, after $8\cdot 10^{7}$ time
steps in a stable configuration. All the barriers are above a critical
threshold, except for those around 1500, where an avalanche is
clearly visible.  (b): Probability distribution of the barriers, $P(B)$, shown
in (a). While the distribution is equal to zero on the left hand side,
a plateau is evident for  $B>B_{c}\sim 0.667$. In the thermodynamic
limit the plateau will become  exactly a constant. In this case the
simulation has been carried out for  $N=5000$ species. (c): 
Probability distribution, $P(T)$, of avalanche duration, $T$,
 in the stationary regime for
the Bak-Sneppen model. The avalanche time series has been recorded in
the stationary state of the system, after $\sim 10^7$ iterations for
$N=2000$ species.  The power law exponent is $\gamma \sim 1$.}
\label{fig1}
\end{figure}

{}From numerical~\cite{Bak93} and analytical~\cite{Flyvbjerg93} studies
it has been shown that the values of the fitness evolve to a step function, 
in the  thermodynamic limit ($N \rightarrow \infty$), characterized by a
single value, $B_c$. For $B<B_c$ the distribution of fitness, $P(B)$,
is uniformly equal to zero while for $B>B_{c}$ we have
$P(B)=1/(1-B_{c})$, determined by the normalization condition. An
example is shown in Fig.~\ref{fig1} (a) and (b).

In this model it is also possible to define a bust-like avalanche
dynamics.  Suppose we fix a threshold for the fitness, $B_{0}$ and
consider  $B_{m}(t)$ as the minimal fitness at time step $t$. If at a
certain  time step, $t_{1}$, it happens that $B_{m}(t_{1})<B_{0}$ then
we can measure the interval of time, $T$, needed for having again
$B_{m}(t_{1}+T)>B_{0}$. In this case an avalanche of duration, or
size, $T$ has taken place in order to restore a minimal fitness in the
system.   If $B_{0}=B_{c}$ then we have $P(T)\sim T^{-\gamma}$: the
system is critical, see Fig.~\ref{fig1}(c).

\begin{figure}
\vspace{1cm} \centerline{\epsfig{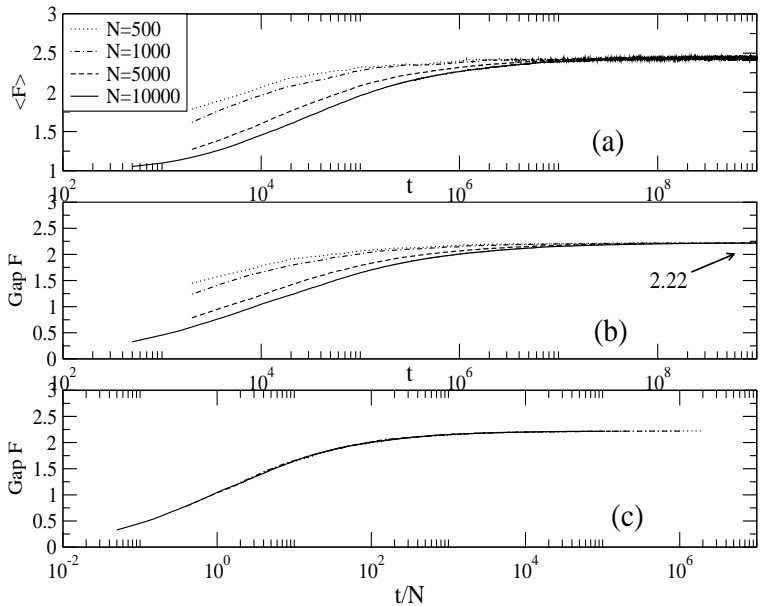}}
\caption{ Average value (a), $\langle \dots \rangle$, and gap (b),
 {\em Gap}, function for the global fitness $F$ and different system size
for simulations up to $10^{9}$ mutations, $t$. 
Note that the gap function 
converges approximately to the value $F_c \sim 2.22$ that corresponds
to the critical threshold for this model. A simple rescaling of the 
time, $t \rightarrow t/N$, brings the curves to collapse into a single
one, as shown in (c).}
\label{fig2}
\end{figure}

According to this model the great mass extinctions of species, like
dinosaurs for example, can be explained in terms of burst-like
dynamics.  A small perturbation in an critical self-connected system
can trigger a chain reaction that may influence a great part of the
species in the ecosystem. Time series of fossils samples seem  to be
in agreement with this avalanche dynamics in the extinction/evolution
of species. A more detailed discussion of the BS model goes beyond
the scope of the present work. For a general review see Ref.~\cite{Paczuski96}.


Despite its simplicity, the BS model evolves
according to a complex dynamics and it is able to explain some 
empirical features of the biological evolution~\cite{Bak93}.
An implicit assumption in the model is that every species
is deeply connected to its environment. A mutation on a single
element automatically triggers a mutation in its neighbors.

But is this approximation always appropriate?  Consider for example
three species in a one dimensional array and suppose  that
$B_{i}=B_{min}$  while $B_{i-1}\cong B_{i+1} \cong 1$.  In the
standard BS model the $i$th cell undergoes to mutation that also
triggers  a change in $B_{i-1}$ and $B_{i+1}$. From a biological point
of view it means that two extremely well adapted species have to
mutate in order to cope with the mutation of the $i$th species.  This
can be interpreted as a very particular (pessimistic) case -- such as,
for example, the case where   the $i$th species is the main source of
food  for both the other species.

In order to stress this idea we use some examples from different areas
in which a similar evolutionary dynamics can be applied.  Suppose that
a new  unfit or unskilled player joins a strong team. Will this player
trigger a regression in the team performance or  will the team
compensate for this lack of skills? This is a small perturbation after
all.

Another example comes from economics. In this case,
it has been shown~\cite{Vandewalle01,Bonanno03,Souma04}, that the
dynamics of different firms is correlated. In fact, it is not unusual for 
a company to own large amounts of stock of other companies and so on.
The result
is an entangled environment, where the evolution of a firm is, in a way, linked
to the evolution of its network of interaction.
Is it then possible, in this case, for an wealthy
environment to sustain an unfit element, or will its lack of ``fitness'' 
bring to the brink of the financial collapse all the other partners,
 as the BS model would suggest?

We provide an answer to these questions using a modified
version of the BS model that takes into account the feedback of the
environment on the single element. We refer to this model incorporating Local
Interactions in the BS model as the LIBS model.  For the sake of simplicity we do
not consider  the topology of the interaction, that may be very
complex; rather, we use a simple 1D array.  The influence of the
network structure on the dynamics of the model will be discussed in
our future work.


As a first approximation we consider our species to be arranged on a
one dimensional array with nearest neighbor interactions. This means
that the micro-environment is composed of three cells.  The value of
the fitness, $B$, for each cell is taken, according to the BS model,
from a uniform distribution between zero and one.  The fitness
parameter, $B_{i}$, of the $i$th cell  represents the {\em
self-fitness} of the species. Motivated by the aforementioned
examples,  we add an environmental  contribution to the self-fitness
that leads to  a {\em global-fitness}, $F_{i}$,  according to
\begin{equation}
F_{i}=B_{i}+\Lambda_{i,i-1} B_{i-1}+\Lambda_{i,i+1} B_{i+1},
\label{global}
\end{equation} 
where $\Lambda_{i,i-1}$ and  $\Lambda_{i,i+1}$ 
are the fractions of fitness that the $i$th cell shares with its neighbours.
The matrix of $\Lambda$s is not symmetric, 
reflecting the fact that the contribution in 
one direction can be very different that the contribution in the
other. This is equivalent to considering a directed {\em weighted graph} with
a trivial necklace topology. 

In the sport example, the global fitness corresponds to the fit players
that contribute to sustaining the unskilled team-mate.
From the economic point of view it represents the capability
of a firm to gain benefits from its partnerships with other firms.
In this particular case, $B_{i}$ represents the wealth generated by the
firm itself, while the other two terms represent the contribution, in different 
forms, from the linked firms. In general, we can consider 
the new terms in the definition of $F_{i}$ as short ranged random
forces acting on the $i$th cell. 

At the beginning of the simulation the self-fitness is drawn from
a uniform distribution between zero and one. The same is done 
for the link weights, $\Lambda_{ij}$. It is worth emphasizing that, in general, 
for two cells $i$ and $j$,  $\Lambda_{ij} \neq \Lambda_{ji}$.

Assuming that the neighbours can cooperate in defining the fitness of a species 
(optimistic view), the extremal dynamics is moved from $B_{min}$
to $F_{min}$. Once the site with minimum global fitness, $k$,
is located, then the self-fitness and the 
interactions of this species 
are redrawn according to the following rules:
\begin{eqnarray}
 \Lambda_{k,k-1}\rightarrow u_{1}, \nonumber \\ 
\Lambda_{k-1,k}\rightarrow u_{2}, \nonumber \\
 B(k) \rightarrow u_{3},\nonumber \\ 
 \Lambda_{k,k+1}\rightarrow u_{4}, \nonumber \\ 
\Lambda_{k+1,k}\rightarrow u_{5},
\label{rules}
\end{eqnarray} 
where the new values for the changed quantities are taken from a
uniform distribution between zero and one,  as in the BS model.
However, in contrast to the BS model,  a change in the fitness of the
$k$th species does not   automatically trigger a change in the
neighbours. Only the interactions are changed.

In order to test the stability of the model we monitor the {\em
average  fitness} and the {\em gap function}, $G(t)$, for both $B$ and
$F$.  The gap function is nothing but the tracking function of the
minimum of $B_{min}(t)$ ( or $F_{min}(t)$). At $t=0$ we have
$G(0)=B_{min}(0)$ ( or $G(0)=F_{min}(0)$). As the evolution proceeds
eventually for a certain $t_{1}$ we will have
$B_{min}(t_{1})>B_{min}(0)$ ( or $F_{min}(t_{1})>F_{min}(0)$) as the
minimum barriers are converging toward the critical value. The gap
function is then updated  as $G(t_{1})=B_{min}(t_{1})$ ( or
$G(t_{1})=F_{min}(t_{1})$) and so on.  It is easy to see that in the
stationary state the gap function converges toward the critical value
\footnote{ For simulations on a finite BS systems, a perfectly stationary
state during finite number of mutations can never be achieved. This
drawback, discussed in Refs.~\cite{Tabelow01,Head02},
 is due to spurious correlations in the  dynamics of the avalanches
induced by the finiteness of the lattice as $G(t)$ gets closer to the critical value
and, therefore, their average duration is suppose to
diverge.  As soon as we get very close to this point, an artifact
regime sets in and the gap function start to saturate toward
$B=1$. The phase in which $ G(t) \sim B_c$ can be regarded as a
transition point for the physically meaningful state:  the larger the
system is, that is closer to the thermodynamics limit, the slower is
the drift  from this point and the system can be regarded, in good
approximation, as stable.   An accurate study of this phenomenon in
relation to the LIBS model,  although very interesting, is not of
fundamental importance in the contest of the present work, therefore
we will consider the system to be stable as  soon as the gap function
and the average reach a plateau.}.

In Fig.~\ref{fig2}  the time series of average values and the gap
function of  $F$ are plotted for  different number of species in the
ecology. The time to reach the stable state depends strongly on the
size: for $N=10^{4}$, the largest system in our simulations, we need
approximately $t \sim 10^{8}$  mutations to achieve the equilibrium.
Note also that a simple rescaling, $t \rightarrow t/N$, leads to a
collapse of these curves. The relaxation times in the BS model 
are, approximately, one order of magnitude lower compared to the LIBS model
of the same size (or in rescaled time).

\begin{figure}
\vspace{1cm}
\centerline{\epsfig{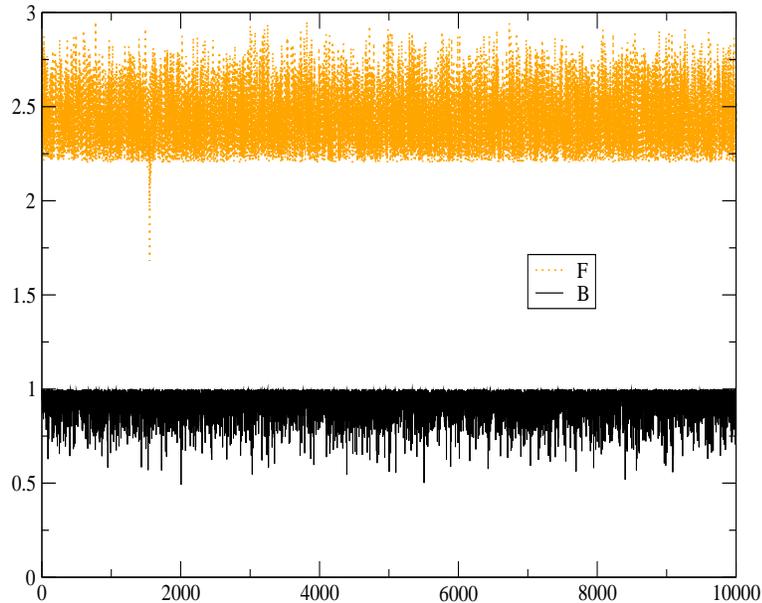}}
\caption{Snapshot of the grid, $N=10^{4}$, in a stable configuration 
for fitness $B$ and global fitness $F$.}
\label{fig3}
\end{figure}

\begin{figure}
\vspace{1cm}
\centerline{\epsfig{figure=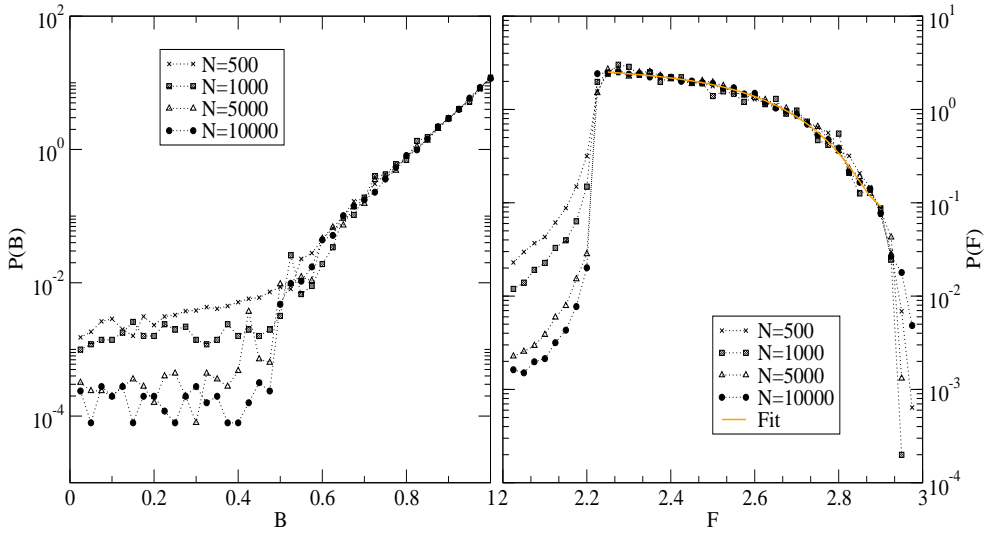,height=7cm, width=13cm}}
\caption{ Left: probability distribution function, $P(B)$, for fitness $B$. 
The step distribution of the standard BS model has been replaced
with an exponential distribution with a cut-off at $F \sim 0.55$.  
Right: Probability distribution, $P(F)$, for the global-fitness, $F$,
and fit with a fourth order polynomial. In
this case a sharp threshold is visible, as in the standard BS model, indicating 
that poor fitness environments undergo mutation. 
 As we consider larger $N$ the transition, at $F_c \sim 2.22$, 
gets sharper and sharper as expected by a finite size analysis.
 The values of the $P(F)$ below this threshold
are related to the recorded avalanches. The distributions shown in these 
plots are the results of an average over 50 different configurations in the
stable regime.}
\label{fig4}
\end{figure}

A snapshot of the grid in the stable configuration is shown 
in Fig.~\ref{fig3}. We notice immediately that that the
local fitness is no longer distributed like a step function
(as for BS). Rather a long, exponential, tail of low fitness is evident, 
as shown in Fig.~\ref{fig4} (Left). The cells with a higher local fitness
still have a greater probability to survive but the global-fitness,
or the presence of environmental partnerships, widens the
possibility of survival, even for some species with a lower degree of 
self-fitness. 

\begin{figure}
\vspace{1cm}
\centerline{\epsfig{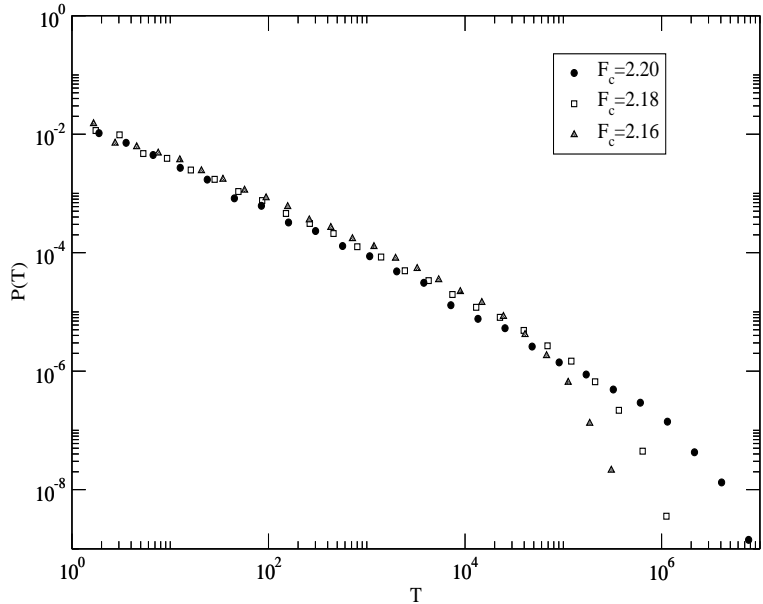}}
\caption{ Probability distribution, $P(T)$, of the avalanche duration, $T$,
in the LIBS model. Three different thresholds
during the stable regime are considered. The critical exponent
is $\gamma \sim 0.83$. For $F_c=2.20$ we are very close to the 
critical point and very large avalanches are present.}
\label{fig5}
\end{figure}

If we examine the global-fitness, a single avalanche is present -- as
in the classical BS model. Moreover the probability distribution
function  for the avalanche duration, shown in Fig.~\ref{fig5} and
computed with respect to $F$, is power law distributed,  in relation
with the criticality of the model.  The index of the distribution
turns out to be different from that of   BS: the change in the
dynamics has also led to a change in the universality class of the
model.

The distribution of global-fitness, shown in Fig.~\ref{fig4} (Right),
has a polynomial decay (4th order fit  in the plot) above a critical threshold,
as a  result of the convolution of stochastic variables. A similar
behaviour can be found also in the size distribution of firms, 
suggesting a possible practical application to the LIBS
model.  Axtell~\cite{axtell01} analyzed the size distribution of U.S. companies, 
defined as the total number of employee, during 1997. He fund that it could be
well represented by a Zipf distribution, $P(s) \sim s^{-\alpha}$ with $\alpha \sim 1$,
being $s$ the size of the firm.  This seemed to confirm
the validity of the ``Gibrat's principle'' according to which the firm
growth rate is independent on its size.  Further investigation on this
issue has been carried on by Gaffeo {\em et al.}~\cite{Gaffeo03},
which analyzed a database of companies for the G7 countries
from 1987 to 2000. Their analysis could confirm the findings of
Axtell, power law distribution with $\alpha \sim 1$, just for some particular
definition of firm size, that is not unambiguous, and for
some particular  business periods. What they found, in general, is a
robust power law behaviour, although the index could change with the
time window analyzed and  the definition of firm size, in contrast with 
the standard theory of Gibrat. 
  The qualitative
discrepancies between $P(F)$ and the distribution empirically found 
for the size distribution of firms can be explained if we
take into account more complex topologies for interactions between
species, or economical entities in this case. Different kinds of
convolution can generate a different shape in the distribution of the
global-fitness  as it can be easily  deduced by writing
Eq.~\ref{global} in a general form as
\begin{equation}
F_{i}=\sum_{j=1}^{k_{i}}\Lambda_{i,j} B_{i,j},
\label{global_gen}
\end{equation} 
 where $\Lambda_{i,i}=1$ and the sum over $j$ is extended to all the
$k_{i}$ neighbours of the $i$th species. In Eq.~\ref{global_gen} no
particular topology is specified.  For an isotropic model on a
$D$-dimensional lattice, $k_{i}$ is equal for all the species and
depends just on $D$ and the definition of neighborhood: the
theoretical boundaries for $F$ are equal for all the
species and we can talk about a ``democratic'' model.  However, recent studies
have shown~\cite{albert02,dorogovtsev2} that, in  real biological and social
systems the number of links per elementary unit, $k$, is not constant,
but characterized by a non-trivial probability distribution function,
$P(k)$, as a result of the complex nature of the interactions between
species or  individuals\footnote{ Two widely studied networks in
literature are {\em random} and {\em scale-free}, which degree is,
respectively, Poisson and power law distributed.  Especially the
latter seem to describe quite well the topology of interactions in
biological, social and technological systems. For a  modern review of
network theory see Refs.~\cite{albert02,dorogovtsev2}.}.  From Eq.~\ref{global_gen},
we can immediately see that, by adopting a complex network as
underlying structure for the interaction between species, we move to
a model in which each species have different boundary values for the
global-fitness, since $0 \leq F_{i} \leq k_{i}$. This inequality have
a straightforward interpretation: species with a large number of
connections will have an higher barrier against  environmental changes
due to the fact that  they can relay on numerous  resources. A simple
way to obtain a complex network structure is   by considering an open
system, where the number of species is not fixed but grows in time, as
for example the firms in a dynamic economy. In this case, it will be
more likely for the new economic entities to be  connected with a well
established one that has already  a large number of connections
(growth and preferential attachment are  actually the two main
ingredients for the Albert-Barab$\acute{a}$si model for  scale-free
networks~\cite{albert02}).  According to our model these ``hubs'',
 that is companies such as General Motors or Coca Cola, have
an higher chance of surviving a turbulent period compared to isolated
nodes: they  have a larger influence in the dynamics of the
model\footnote{This situation of ``freezing'' of the large ``hubs''
has some analogies with spin glass theory where some species freeze in
a random  configuration leading to a rough landscape of energy levels
at small  temperatures~\cite{parisi}.}.  This simple consideration,
although not exhaustive, show how the underling topology of the
interactions can play an important role in the final distribution of
the global fitness. Further analytical and  numerical test would be of
great importance in order to understand the dynamics of the LIBS model
and to which  extend it can be applied to real systems. 
It is also worth pointing out that
another parameter related to $P(F)$ is the
domain of $B$ itself, that, in the present, case we assume to be
uniform in the interval (0,1). In fact, a change in this distribution,
while preserving the dynamics of the model, would lead to a different
shape in the final distribution of $F$. 
 While these important issues will be addressed in our future work, in 
Appendix  we report a further extension of the model where also
the second-nearest neighbour interaction is considered.

The results obtained with the LIBS model confirm the relevance
of self-organized criticality in complex systems and, in particular,
economics. The concept of mutual cooperation, introduced via the
global-fitness, can explain the ubiquity of broad tails in the
distribution of characteristic quantities of physical and social
systems in terms of a convolution of variables between elements of the
network of interaction.  In the economic context, this asymptotic
behaviour can be related to the empirical findings  concerning the
distribution of the size of firms.   The possible relevance of
self-organized criticality in economics has already been suggested by
recent theoretical and empirical
studies~\cite{Bak93b,Bak97,ponzi00,bartolozzi04},  while possible
applications of the BS model in this field can be found in
Refs.~\cite{cuniberti01,yamano01,ausloos04}.  The application of the
SOC concept to social sciences can, in general, be motivated by
empirical observations of the ``intermittent'' activity in the human
dynamics at every level, from wars to revolutions and, in particular,
intellectual production where moments of frenetic activity can be
alternated by long breaks, which length cannot be predicted (this
holds, indeed, for one of the authors, M.B.).  This process is, in a
way, similar to the discharge, via avalanches, needed in the classical
sandpile model, to restore the critical slope.  In a real
economic world a wide series of changes, similar to avalanches, can
be  triggered by exogenous or endogenous shocks related to
structural changes at macroeconomic level, for example the creation
and successive enlargement  of the European community or the fall  of
the Soviet empire, or at microeconomic level, as  the invention of a
new technology~\cite{day_book}.   Since the shocks leading to
avalanches are of different nature, we also  expect the existence
of different time scales involved in the  self-organization process.
In SOC systems, in general, the existence of a
sharp separation between time scales, energy storage and relaxation, appears to
be a strict prerequisite.   In the BS model, as in the LIBS
model, by mutating  one unit at the time, we implicitly assume that
the time to extinction, $t_e$, of a species depends exponentially on
its global fitness,  that is $t_{e}(i)\propto e^{F_i}$. The
exponential separation of the extinction times is at the core of the
``punctuation''. In economics terms  we can still assume this
behaviour: changes of poorly fitted firms can be simply related to
small microeconomic fluctuations that can happen in time scales of the
order of weeks or months while much longer times are needed to change
the  fitness of an highly adapted company. In the latter case 
radical changes  are needed, as for example a switch from a political
regime to another, that may take centuries to happen.



In conclusion, we have extended the Bak-Sneppen model for
biological evolution  by introducing explicitly local
interactions between elements of the ecology.  Numerical simulations
have shown how the dynamics of the model, while still leading to  a
self-organized critical state, can be largely effected by the
environmental forces, leading to smoother distributions in both the
intrinsic fitness, $B$, and the global fitness, $F$.  As already
pointed out by Grassberger~\cite{grassberger95} the BS model cannot 
be taken too seriously for describing the
punctuated equilibrium of biological evolution. Nevertheless,
because of its simplicity, it can easily be used as paradigm for other
complex systems. In the present work we suggest a possible application
of our extension of the BS model to the economic world.  In particular
the distribution of global-fitness can be related to the size
distribution of firms in the most developed markets.  In this respect
the evolution of firms is seen as a punctuated equilibrium process in
which the convolution of mutual interest can justify the spreading in
size of the firms themselves.  It is worth pointing out
that the actual shape of the distribution of global-fitness is related
to the topology of the interaction.  A simple 1D model cannot 
be expected to account completely  
for the power law distribution observed in the size of
firms. Future work will be devoted to the application of the  
model proposed in this
paper to more complex topologies -- as, for example, to
scale-free networks~\cite{albert02,dorogovtsev2},
that are more likely to reproduce the real interactions between
economic entities.

\section*{Appendix}

We consider now a further extension of the LIBS model by including
 the second nearest neighbours interactions in the simple 
one dimensional topology. The global-fitness of Eq.~\ref{global} becomes
\begin{equation}
F_{i}=B_{i}+\Lambda_{i,i-1} B_{i-1}+\Lambda_{i,i+1} B_{i+1}
+\Lambda_{i,i-2} B_{i-2}+\Lambda_{i,i+2} B_{i+2},
\label{libs_2n}
\end{equation} 
where we the second order coefficients are not independent
random numbers but  $\Lambda_{i,i \pm 2}=\Lambda_{i,i \pm 1} \cdot
\Lambda_{i \pm 1,i \pm 2}$.  The reason behind this choice, that can
be easily extended to the $n$th order neighbours, is motivated by the
assumption that the higher order interactions are  damped by the
``distance'' between the two species and therefore $\Lambda_{i,i \pm 2} \leq
\Lambda_{i,i \pm 1}$.
  By using this formulation, we attempt to mimic a hierarchical
dependence of the global fitness in the ecology: species become
explicitly related to their second nearest neighbours via the mediation
of  their first neighbours and so on.  Using these constrains
 the mutation rules remain the same as in
Eq.~\ref{rules} since a change in the first order coefficients
triggers automatically a change also in the higher order ones.

\begin{figure}
\vspace{1cm}
\centerline{\epsfig{figure=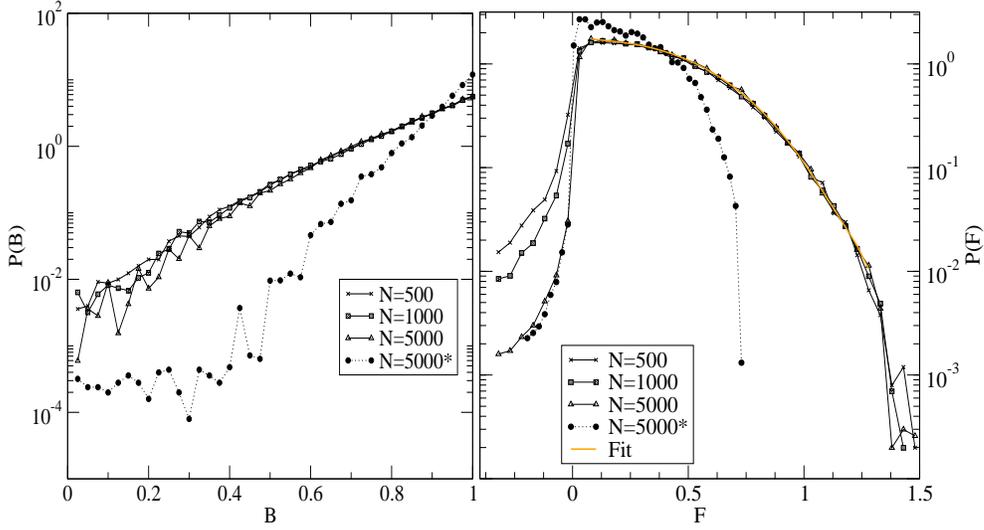,height=7cm, width=13cm}}
\caption{ Left: distribution of self-fitness, $B$, for the second 
order neighbours and different grid size, $N$. The distribution for the
first order neighbours is also plotted (*). Note that an extension of the
neighborhood leads to a slower decay rate. Moreover, a  clear 
cut-off for low $B$ is not evident anymore. Right: distribution of global 
fitness $F$. In this case the distributions have been rescaled according to
$F \rightarrow F-F_c$. For the second order neighbours we have $F_c \sim 2.37$.
The fit with a sixth order polynomial, also shown, produces a smaller 
$\chi^{2}$ compared to the fourth order used for the first nearest neighbour case.} 
\label{fig6}
\end{figure}

The dynamics that results from the numerical simulations is similar to
the  first neighbours LIBS model: after an extensive transient we
reach a critical  stationary state characterized by avalanches of
mutations, which size, $T$, is power law distributed.  The
distributions for $B$ and $F$, in the stable regime, are shown in
Fig.~\ref{fig6} along with the  distribution obtained by considering
just the first neighbour interaction.  In the case of $B$ we can
notice that by enlarging the neighborhood the  distribution show a
slower decaying rate and they  appear to be smoother. In this case we
have an exponential decay all the way down to zero, without any clear
cut-off for low $B$. Despite their fitness, all the species have a
chance of survival if sustained by an healthy environment.  Regarding
the global fitness, instead, a polynomial decay is still evident,
although the order is higher compared to the first neighbours case. In
is also important to notice that a relative large change in the
theoretical range for $F$, which bounds are now $0 \leq F_{i} \leq 5$,
does not lead to a consequent rise in the threshold value that moves
just slightly from $F_c \sim 2.22$ to $F_c \sim 2.37$.  This is
equivalent to say that, in the previous case, in order to be
considered ``fit'',  a species has to exceed the $\sim$ 74\% of the
possible range for $F$, now the $\sim$ 48\% is sufficient!  In
conclusion, a hierarchical extension  of the cooperation between
species in the LIBS model  leads to an easier adaptation and survival
probability: the more compact the ecosystem is the higher will be the
chances of survival of each single species as long as they cooperate
for their mutual interest.

\section*{Acknowledgement}
This work was supported by the Australian Research Council.

\end{document}